**Response to comment on "Mutualisms weaken the latitudinal diversity gradient among oceanic islands"**

*Camille Delavaux, Thomas Crowther, James Bever, and Evan Gora*

In our original paper (Delavaux et al. 2024; https://www.nature.com/articles/s41586-024-07110-y), we find that the latitudinal diversity gradient (LDG) in plant species richness is reduced on oceanic islands worldwide. Moreover, we find that plants that associate with mutualists, including pollinators, AM fungi, and N-fixing bacteria, disproportionately contribute to this reduction. To make this assertion, we fit models of continental diversity to absolute latitude and compared mainland diversity estimated by these models to the observed diversity of individual islands. The biological interpretation of a mutualist filter (i.e. reduced colonization from mutualist associated plant species) from these analyses are consistent with several independent analyses (Simonsen et al. 2017, Delavaux et al. 2019, Taylor et al. 2019, Delavaux et al. 2021, König et al. 2021, Delavaux et al. 2022) that do not use information on latitude. Thus, these patterns are robust to different statistical approaches.

The critique of our work by Pitcher and Hartig (https://arxiv.org/pdf/2411.15105) questions our results through a reanalysis of the data. In their reanalysis, they add longitude as a predictor of richness on islands. However, we have several concerns about this reanalysis. In particular, we question (1) the biological/physical basis of predicting species richness with longitude as well as modelling non-linear relationships potentially leading to spurious relationships and (2) the justification for how this comment calculated island deficits from the data. Additionally, (3) described analyses and provided code (https://github.com/MaximilianPi/IslandLDG) are in many cases insufficient to understand or meaningfully engage with this work, precluding productive discussion about cited analytical concerns. Given these major issues with the reanalysis, we reject the conclusions from this critique that our prior results are due to statistical artifacts. We therefore do not have reason to question our original conclusion that *the latitudinal diversity gradient is weakened by mutualisms among oceanic islands worldwide.*

1.   **Physical and biological expectations**
     **A. Lack of biological/physical basis for the use of longitude to predict species richness**
     *Latitude is a meaningful predictor of species richness*
     Latitude determines patterns of solar incidence which is a major driver of climate, including temperature and precipitation. As a biological response to this climate gradient, there is an obvious and consistent gradient in plant diversity with latitude that has been the focus of inquiry from biologists for decades. This same gradient of physical factors should also influence the species diversity gradient on islands, thereby providing justification for our approach. We acknowledge that there is substantial noise around our fit of diversity and composition with latitude. However, that is the nature of the data.

     *Longitude is not a meaningful predictor of species richness*
     Pitcher and Hartig provide no biological reason that longitude would be a valid predictor of species richness. We are not aware of a biological/physical basis for predicting species richness or mutualist patterns with longitude. Longitude is perpendicular to latitude and does not correlate with any physical difference at the global scale. There are interesting longitudinal

climate and biological gradients determined by proximity to mountains or oceans, but these are regional patterns, not global patterns. The influence of mountains or oceans on climate does not arrange itself in a coherent manner with longitude. As a result, there are regions in which increasing longitude is correlated with an increase in precipitation and regions in which an increase in longitude is correlated with a decrease in precipitation. At the global scale, there are no consistent or obvious longitudinal gradients in climate or biological diversity. Fits for longitude will then likely be driven by spurious associations with other drivers. We also note that longitude is a biased measure of distance. While the difference in latitude at a particular longitude is consistently spaced, the difference in longitude varies with latitude, going from 69.2 miles per degree longitude at the equator to zero miles per degree at the poles. This bias itself could generate spurious covariance with latitude and distance.

**B. Potential spurious relationships by building models that do not have a physical or biological basis**

*Inclusion of longitude into global models of biological diversity or composition is not justified*

When metrics are included in analyses without prior motivation, it is well known that patterns discovered are likely to be spurious. Authors of this comment appear to be extrapolating from longitudinal patterns on continents to project to islands. This is a very questionable activity, not just because it projects a potentially spurious correlation, but because of the nature of longitudes. As we state above, longitudes do not consistently map onto distance. Moreover, continents cover only 29% of the earth and, while they do a fair job of representing latitudes, they are very unequally distributed across longitudes. Hence, the authors are left fitting data to very limited representation of longitudes and then projecting a great distance at the equator (e.g. to islands in the Pacific), shorter distances close to the poles. Therefore, there is likely an unavoidable longitudinal gradient in error of the projection of a potentially spurious pattern. It is therefore not a surprise that this reanalysis finds a weaker relationship.

*Non-linear relationships could be consistent with mutualism interactions*

It is not clear why the mutualism filter would not be a plausible explanation for the patterns observed in the second approach undertaken by Pitcher and Hartig. Specifically, it is possible that the non-linear approach used for area, distance, and latitude could capture the effect of the mutualism filter strength. We explicitly tested for interactions between mutualism filter strength and all other variables, including latitude, and found they were significant. Interactions between two linear variables can alternatively be fit as a non-linear relationship between one of those variables and the response. Additionally, it is not clear why Pitcher and Hartig treated all variables as non-linear except the mutualist filter, as no justification is given for this different treatment.

2. **Assigning expected mainland richness to islands**

It is unclear how Pitcher and Hartig justify assigning expected species richness to a given island using both latitude and longitude. If the authors are justifying inclusion of longitude based on likely origin of plant seed, this would be a very complicated calculation which would rely on major assumptions about dispersal patterns. Because the authors do not provide information on this approach and declined to provide this information upon receiving our initial response to this comment, we cannot evaluate the results comprehensively. We do not see a way that this could be done appropriately unless dispersal pathways are explicitly modeled or expected island richness is determined by averaging across the great number of potential source environments, which would be functionally similar to our latitudinal approach. We expect major skepticism will remain unless these assumptions are explicitly detailed and the sensitivity of the results to those assumptions is evaluated.

Moreover, the machine learning approach used by Pitcher and Hartig is an excellent option for producing a better fit to the training data, but misuse of machine learning often produces overfitting to the data and spurious relationships. In this particular context, we expect this machine learning approach to produce flawed predictions because the training data are unevenly distributed across latitudes and longitudes, confounding their effects. To test whether this approach is predicting unrealistic expectations of 'mainland' diversity - a key input in the species deficit calculations - we assessed the global species richness values predicted based on the authors' random forest model (code available here: https://github.com/c383d893/IslandLDG-ReReAnalysis under 2_Analyses/ReReAnalysisMap.R). We found that mainland predictions (across mainlands and islands) naively using latitude and longitude in this way produce errors when compared to known plant diversity patterns (Figure 1). For example, known high diversity tropical regions in the mainland (e.g. Ecuador) and islands in the Pacific are predicted by this model to have very low diversity, likely because of the low representation of low-latitude data at these longitudes. The fine-grain resolution of their model also appears to predict major differences in diversity at fine-scales that are likely due to the sparse and variable nature of the underlying data rather than consistent fine-scale differences in richness. Moreover, this model is meant to predict the richness of the mainland sources for each island location as an estimate of the richness than a 'mainland' could have at that location, but it is apparent from the predictions (Figure 1) that this model is failing in that goal because the predicted richness over island locations differs strongly from their nearby source locations for most oceanic locations. In conclusion, the higher $R^2$ of the random forest model of Pitcher and Hartig is misleading because a simple visualization of these trends clearly demonstrates that this is a classic case of overfitting limited training data to produce incorrect and spurious results.

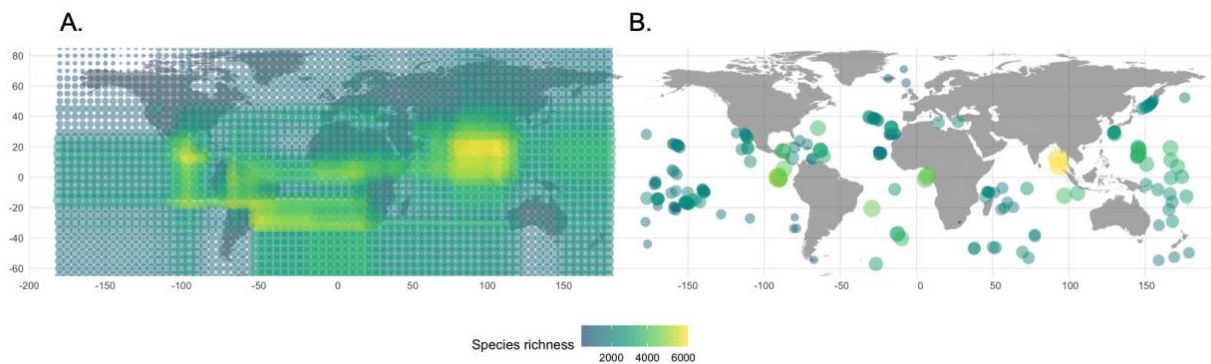

**Figure 1.** A global projection of predictions based on Pitcher and Hartig's random forest model including latitude and longitude (A) across a grid of world locations and (B) across islands used in Delavaux et al. 2024. Some regional predictions underpredict known values, raising concerns about the validity of this approach.

Overall, we found ourselves needing to make several assumptions about the analyses and figures, making it difficult to give informed comments. We found many other small errors or unexplained inconsistencies that made interpretation very challenging (mislabeled axes, unclear methods, etc.). We also note that in an attempt to engage with the reanalysis, we downloaded the github repository provided by the authors of the critique and it was not fully repeatable and poorly annotated. Here, we highlight two examples that demonstrate the lack of clarity related to the reanalysis.

First, Figure 1 A and B from Pitcher and Hartig are based only on AM filter strength. We understand that this is an example of the difference in predictions between the original generalized additive model (GAM) and random forest (RF) approach. However, it is not possible to understand whether Figure 1 C and D are testing altered results from these predictions only (AM filter) or across the mutualism filter. If they are only for the AM filter, this is not comparable to the mutualism filter (across three mutualisms) tested in our study. It is therefore not possible to evaluate this approach relative to the original peer-reviewed analysis.

Second, it is unclear to us what the authors aim to show with Figure 3C. This is referenced only in the last sentence of the critique, claiming that "The lack of a systematic differences between mutualism strength on islands and the corresponding mainland locations (Fig. 2C) is adding to our doubts if a strong effect of mutualists on island species richness truly exists." The mutualist filter strength on mainlands represents the proportion of plant species interacting with one or more mutualists. We indeed hypothesize and show in our paper that where this value is higher, this will lead to a more acute loss of species on islands (Delavaux et al. 2025, Figure 2C), and resulting higher mutualist contribution to species deficit (Delavaux et al. 2025 Figure 3). Moreover, we show that islands support fewer mutualist taxa than equal-latitude mainlands (Figure 3, Delavaux et al. 2024), as do many prior analyses (Delavaux et al. 2019, Taylor et al. 2019, Delavaux et al. 2021, König et al. 2021, Delavaux et al. 2022). We show these trends in our study and therefore find support for the latitudinal structuring of species deficit on islands.

In conclusion, there are major reasons to be skeptical with interpretations from the reanalysis approach used by Pitcher and Hartig. The major concerns that remain are the lack of (1) the biological/physical basis of predicting species richness with longitude as well as modelling non-linear relationships potentially leading to spurious relationships and (2) the justification for how this comment calculated island deficits from the data. Moreover, (3) the lack of clarity in the comment's text, figures, and associated code makes evaluating this critique very challenging, and not conducive to healthy debate about the relevant ecological processes. Given these major flaws and the abundant evidence that the mutualism filter exists (as supported by independent analyses: Simonsen et al. 2017, Delavaux et al. 2019, Taylor et al. 2019, Delavaux et al. 2021, König et al. 2021, Delavaux et al. 2022), we conclude that their re-analysis does not undermine our original conclusion that the ***mutualism filter weakens the latitudinal diversity gradient among oceanic islands worldwide.***